\documentclass[prl,superscriptaddress,amsmath,amssymb,twocolumn]{revtex4-2}

\usepackage{graphicx}
\usepackage{bm}
\usepackage[colorlinks = true]{hyperref}
\usepackage{physics}
\usepackage{upgreek}
\usepackage{xcolor}

\usepackage{CJKutf8} 

\usepackage{mathtools}
\mathtoolsset{showonlyrefs, showmanualtags}

\begin{document}

\title{Exact bound of power-efficiency trade-off in finite-time thermodynamic cycles}

\author{R. X. Zhai (\begin{CJK}{UTF8}{gbsn}翟若迅\end{CJK})}
\affiliation{Graduate School of China Academy of Engineering Physics, Beijing 100193, China}

\author{Xin Yue (\begin{CJK}{UTF8}{gbsn}岳鑫\end{CJK})}
\affiliation{Beijing Computational Science Research Center, Beijing 100193, China}
\affiliation{Graduate School of China Academy of Engineering Physics, Beijing 100193, China}

\author{C. P. Sun (\begin{CJK}{UTF8}{gbsn}孙昌璞\end{CJK})}
\email{suncp@gscaep.ac.cn}
\affiliation{Graduate School of China Academy of Engineering Physics, Beijing 100193, China}

\date{\today}

\begin{abstract}
    Power and efficiency are fundamental criteria for evaluating the performance of thermodynamic cycles. However, it is generally impossible to maximize both simultaneously. In particular, achieving maximum efficiency inevitably leads to vanishing power as the cycle duration approaches infinity. A quantitative characterization of this trade-off yields significant theoretical and practical implications. In this letter, we analytically derive an exact bound constraining power and efficiency in low-dissipation finite-time heat engines. This bound specifies the maximum power attainable at any prescribed efficiency, thereby providing a benchmarking for evaluating the performance of heat engines.
\end{abstract}

\maketitle

\emph{Introduction} -- 
The efficiency of a heat engine operating between two reservoirs is fundamentally bounded by the Carnot efficiency \cite{Huang1987,Ma2004}. However, achieving such an upper bound in a macroscopic heat engine requires its thermodynamic cycle to operate in a quasi-static manner, taking an infinite amount of time. In principle, such infinite cycle time results in a vanishing output power, making an ideal engine impractical \cite{Andresen1984,Tu2012}. Gaining finite power unavoidably brings in irreversibilities that cut down the efficiency, so a realistic heat engine inevitably runs below the Carnot limit. This trade-off between power and efficiency has inspired the flourishing study of finite-time thermodynamic cycles. Many of these studies aim to find the maximum power output and determine the corresponding efficiency at maximum power \cite{Yvon1955,Chambadal1957,Novikov1958,Curzon1975,VandenBroeck2005,Sheng2015,Brandner2015,Schmiedl2007,Tu2014,Izumida2008,Tu2008,Allahverdyan2008,Rutten2009,Wang2012,Martinez2015,Rossnagel2016,Esposito2010,Abiuso2020,Liang2025}. A direct generalization on such question is determining the efficiency constrained by a fixed power \cite{Ryabov2016,Holubec2016,Long2016,Shiraishi2016,Cavina2017,Holubec2017,Ma2018,Josefsson2018,Josefsson2019,Chen2022,Zhao2022,Zhai2023,Zhai2024,Ma2024,Ma2024a,Zhao2025,Ye2025}, or vice versa.

To quantitatively evaluate the irreversibility during finite time processes, a rather simple and straightforward assumption is frequently used. Namely, the irreversible entropy production is inversely proportional to the process time \cite{Esposito2010,Zhai2023}, which is usually dubbed as \emph{low-dissipation} assumption. With such \(1/\tau\) entropy production model, one can express the engine's power and efficiency as functions of the time of the processes. Specifically, in Ref. \cite{Esposito2010} the model is initially introduced and the bounds of efficiency at maximum power is derived; In Refs. \cite{Ryabov2016,Holubec2016} these results are directly extended, and the upper and lower bounds are suggested on efficiency for arbitrarily given power; Ref. \cite{Ma2018} presents a tighter bound, under which the conjecture proposed in Refs. \cite{Ryabov2016,Holubec2016}—previously validated only near the extrema of power—is rigorously proven for arbitrary power and efficiency.

Despite such efforts have been dedicated within the \(1/\tau\) entropy production model, the tightest bound between power and efficiency have not been founded exactly. In this letter, we achieve an exact bound for the constraint between power and efficiency. With this bound, the maximum power for arbitrarily given efficiency is obtained.

\emph{Finite-time heat engines} -- 
Consider a heat engine operating between two reservoirs at different temperatures. Its thermodynamic cycle consists of two finite-time isothermal processes and two adiabatic processes. During the two finite-time isothermal processes, the engine exchanges heat \(Q_c < 0\) with the low-temperature reservoir at a temperature \(T_c\) within a time period \(\tau_c\), and \(Q_h > 0\) with the high-temperature reservoir at a temperature \(T_h\) within a time period \(\tau_h\). Here we employ the approximation where irreversible entropy production is inversely proportional to the process time \cite{Ma2020}, giving
\begin{equation}
    Q_{c,h} = \pm T_{c,h} \Delta S + \frac{M_{c,h}}{\tau_{c,h}},
    \label{eq:one_over_tau_approximation}
\end{equation}
where \(\Delta S\) is the reversible entropy change of the working substance during a quasi-static Carnot cycle, and \(M_{c,h} \ge 0\) are the inverse proportional coefficients of entropy production. Therefore, the irreversible entropy productions are always positive. The positive or negative sign in Eq. \eqref{eq:one_over_tau_approximation} is for \(Q_{c}\) and \(Q_h\), respectively.

We consider the circumstances where the time of the two adiabatic processes is much shorter than \(\tau_h\) and \(\tau_c\) \cite{Ma2020,Zhai2023}, so that the total time of the cycle is \(\tau_{\mathrm{cycle}} = \tau_c+\tau_h\). By defining \(m_c = \sqrt{M_c/ T_h \Delta S}\) and \(m_h = \sqrt{M_h /T_h \Delta S}\), the power \(P\) and efficiency \(\eta\) for the finite-time heat engine are expressed as
\begin{align}
    P = \frac{\eta_C - m_c^2 /\tau_c - m_h^2 /\tau_h }{\tau_c + \tau_h} T_h \Delta S,
    \label{eq:def_power}
    \\
    \eta = \frac{\eta_C - m_c^2 /\tau_c - m_h^2/\tau_h}{ 1 - m_h^2 /\tau_h}.
    \label{eq:def_efficiency}
\end{align}
It follows from Eqs. \eqref{eq:def_power} and \eqref{eq:def_efficiency} that as the time approaches infinite, the efficiency approaches the Carnot efficiency \(\eta_C = 1 - T_c/T_h\), and the power reduces to zero, which is in agreement with the former qualitative description of finite-time engines.

The maximum power is obtained at the process times \(\tau_{c,h}^* = 2 (m_c + m_h) m_{c,h}/\eta_C \) as 
\begin{equation}
    P_{\max} = \frac{\eta_C^2 T_h \Delta S}{ 4 (m_c + m_h)^2 }.
     \label{eq:Pmax}
\end{equation}
With this maximum power, a normalized power is defined as \(\tilde{P} = P/P_{\max}\). We re-express Eqs. \eqref{eq:def_power} and \eqref{eq:def_efficiency} into compact form as
\begin{equation}
    \tilde{P} = \frac{4}{\eta_C^2} \frac{(\eta_C - \sigma_c - \sigma_h) \sigma_c \sigma_h }{\xi^2 \sigma_c + (1 - \xi)^2 \sigma_h}
    ,
    \label{eq:sigma_normalized_power}
\end{equation}
and
\begin{equation}
    \eta = \frac{\eta_C - \sigma_c - \sigma_h}{1 - \sigma_h}.
    \label{eq:sigma_efficiency}
\end{equation}
Here \(\sigma_{c,h} \equiv m_{c,h}^2/ \tau_{c,h}\) are dimensionless variables to be optimized, being proportional to the entropy production, and \(\xi \equiv m_h/(m_c + m_h)\) is a dimensionless parameter.

\emph{Maximum power for given efficiency} -- We seek the maximum power \(\tilde{P}_m\) for fixed efficiency \(\eta\). It follows from Eq. \eqref{eq:sigma_efficiency} that \(\sigma_c\) Eq. \eqref{eq:sigma_normalized_power} is eliminated, and the normalized power is obtained as a function of \(\sigma_h\)
\begin{align}
    \tilde{P} (\sigma_h) = 
    \frac{\eta ( 1 - \eta)}{\eta_C^2/4}
    \frac{\sigma_h (\sigma_m - \sigma_h) (1 - \sigma_h)}{\Xi (\sigma_\xi - \sigma_h)},
    \label{eq:P_tilde_single_func}
\end{align}
where \(\Xi \equiv (1 - \eta)\xi^2 - (1-\xi)^2\), \(  \sigma_m \equiv \flatfrac{(\eta_C - \eta)}{(1 - \eta) } \) and 
\begin{align}
    \sigma_\xi \equiv \frac{1}{1 - (1-\eta)^{-1} \qty[(1-\xi)/\xi]^2} \sigma_m
\end{align}
To ensure positive entropy production, i.e., \(\sigma_{c,h} > 0\), the range of \(\sigma_h\) is determined as  \(0 < \sigma_h < \sigma_m\). It can be proved that \(\tilde{P}(\sigma_h)\) has only one maximum point \(\sigma_h^*\) within this domain, which is determined by \(\partial \tilde{P} / \partial \sigma_h = 0\). The maximum power with fixed efficiency is then given with \(\tilde{P}_m = \tilde{P}(\sigma_h^*)\).

Eq. \eqref{eq:P_tilde_single_func} can be rewritten as a ratio \(\tilde{P} = q_1(\sigma_h) / q_2(\sigma_h)\), with
\begin{align}
    q_1 \equiv \sigma_h (\sigma_m - \sigma_h) ( 1 - \sigma_h), \quad
    q_2 \equiv \frac{\Xi \eta_C^2( \sigma_\xi - \sigma_h )}{4 \eta ( 1 - \eta )}.
\end{align}
The conditions of maximum point read as
\begin{equation}
    \eval{\tilde{P}}_{\sigma_h^*} = \frac{q_1(\sigma_h^*)}{q_2 ( \sigma_h^*)} = \tilde{P}_m,
    \label{eq:P_m_func}
\end{equation}
and
\begin{equation}
    \eval{\dv{\tilde{P}}{\sigma_h}}_{\sigma_h^*} = 
    \frac{1}{q_2} \eval{\qty(\dv{q_1}{\sigma_h}
    -
    \frac{q_1}{q_2}
    \dv{q_2}{\sigma_h})}_{\sigma_h^*} 
    = 0.
    \label{eq:P_m_derivitive}
\end{equation}

We construct a cubic function \(f(\sigma_h)\) as
\begin{align}
    f(\sigma_h) & \equiv q_1 (\sigma_h) - \tilde{P}_m q_2 (\sigma_h)  \\
    & = \sigma_h^3 - (1 + \sigma_m) \sigma_h^2
    + 
    \qty( \sigma_m + \Pi) \sigma_h 
    -  \Pi \sigma_\xi,
    \label{eq:Delta_eq_0}
\end{align}
where \(\Pi = \Xi \tilde{P}_m \eta_C^2 / [4 \eta (1 - \eta)]\) is the only parameter among all that related to \(\tilde{P}_m\).
The condition for maximum point in Eqs. \eqref{eq:P_m_func}-\eqref{eq:P_m_derivitive} are re-expressed by \(f(\sigma_h)\) as \(f(\sigma_h^*) = f'(\sigma_h^*) = 0\). This sufficiently guarantees that the maximum point \(\sigma_h^*\) is a repeated root of equation \(f(\sigma_h) = 0\). According to the property of cubic equations, the existence of repeated root sufficiently gives that the discriminant of \(f(\sigma_h)\) vanishes, giving
\begin{align}
    \begin{split}
         & \Delta \equiv (\Pi +\sigma_m)^2 (\sigma_m +1)^2
        - 4 (\Pi +\sigma_m )^3
        - 27 \Pi^2 \sigma_\xi^2 \\
        & - 4 \Pi \sigma_\xi (\sigma_m + 1 )^3
        + 18 \Pi \sigma_\xi ( \Pi +\sigma_m ) (\sigma_m + 1 ) = 0.
    \end{split}
    \label{eq:function_Delta}
\end{align}
Solving \(\Pi\) in Eq. \eqref{eq:function_Delta} yields the maximum power \( \tilde{P}_m\) where the efficiency is constrained to \(\eta\).

\emph{Finding the proper solution} -- Notably, Eq. \eqref{eq:function_Delta} leads to a cubic equation for \(\Pi\) (or \(\tilde{P}_m\)), which can have up to three real roots. To determine the unique maximum power, we should consider further constraint on \(\sigma_h^*\).
By noting that \(f(0) = - \Pi \sigma_\xi < 0\), \(f(\sigma_m) = - \Pi (\sigma_\xi - \sigma_m) < 0\) and \(f'(\sigma_h^*) = 0\), it is further determined that \(\sigma_h^*\) is a local maximum point of \(f(\sigma_h^*)\). Thus, we express \(\sigma_h^*\) with the smaller root of the quadratic equation \(f'(\sigma_h^*) = 0\). Namely,
\begin{equation}
    \sigma_h^* = \frac{1}{3}\qty[
        1 + \sigma_m
        - 
        \sqrt{(1+\sigma_m)^2 - 3(\sigma_m + \Pi)}
    ].
\end{equation}
To ensure \(\sigma_m\) being in the domain \(0 < \sigma_h^* < \sigma_m\), constraints are shown in TABLE. \ref{tab:Pi_constraint}, with which only one solution of Eq. \eqref{eq:function_Delta} is determined as the maximum power with a constrained efficiency, i.e., \(\tilde{P}_m (\eta)\).
\begin{table}[!hbtp]
\begin{center}
    \caption{Constraint to \(\Pi\) posted by the range of \(\sigma_h^*\).}
    \label{tab:Pi_constraint}
    \includegraphics[scale=0.9]{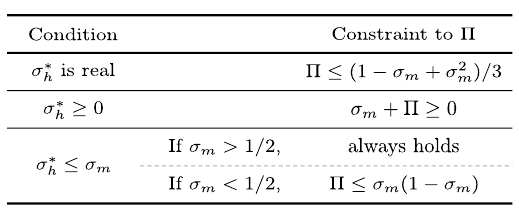}
\end{center}
\end{table}

\emph{Examples} --
The exact constraint bound presented in Eq. \eqref{eq:function_Delta} and TABLE \ref{tab:Pi_constraint} is given in implicit form. In the following discussions, we demonstrate how to obtain an explicit expression for this bound by working through several examples with specified parameters. For certain parameter choices, our formulation reproduces previously reported results, thereby partially validating the correctness of our approach.

(1) For \(\xi \rightarrow 0\) or \(m_c \ll m_h\), we have \(\Xi = - 1 \). Namely, the irreversibility is dominantly contributed by the high-temperature finite-time isothermal process, and Eq. \eqref{eq:function_Delta} is reduced into
\begin{equation}
    \Delta \rightarrow (\Pi + \sigma_m)^2 \qty[ (1 - \sigma_m)^2 - 4 \Pi ] = 0.
\end{equation}
Considering the constraints in TABLE \ref{tab:Pi_constraint}, we obtain
\begin{equation}
    \tilde{P}_m = \frac{4}{\eta_C^2} \eta ( \eta_C - \eta).
    \label{eq:constraint_xi_1}
\end{equation}
This bound is previously obtained as the universal lower limit of efficiency for arbitrarily given power \cite{Ryabov2016,Holubec2016,Shiraishi2016,Ma2018,Zhao2022,Ma2024}.

(2) For \(\xi \rightarrow 1\) or \(m_c \gg m_h\), we have \(\Xi = 1 - \eta\), the irreversibility is contributed by the low temperature finite-time isothermal process, and Eq. \eqref{eq:function_Delta} is reduced into
\begin{equation}
    \Delta \rightarrow (1 - 4 \Pi) \qty[\sigma_m (1 - \sigma_m) - \Pi]^2 = 0.
    \label{eq:Delta_xi_1}
\end{equation}
Eq. \eqref{eq:Delta_xi_1} admits two different solutions, i.e., \(\Pi_1 = 1/4\), and \(\Pi_2 = \sigma_m (1-\sigma_m)\), with \(\Pi_1 > \Pi_2\). The first and second conditions in TABLE \ref{tab:Pi_constraint} are automatically satisfied.
For the third condition, if \(\sigma_m < 1/2\) (equivalently, \(\eta > 2\eta_C - 1\)), only \(\Pi_2\) is admissible; if \(\sigma_m > 1/2\), then both \(\Pi_1\) and \(\Pi_2\) satisfy TABLE \ref{tab:Pi_constraint}, and the maximal power is determined by the larger solution \(\Pi_1\).
Therefore, the constraint relation for \(\xi \rightarrow 0 \) is finally determined as a piecewise function,
\begin{equation}
    \tilde{P}_m = 
    \begin{cases}
        \eta/\eta_C^2 , & \text{for } \eta < 2 \eta_C - 1, \\ \\
        \displaystyle \frac{4}{\eta_C^2} \frac{\eta(\eta_C - \eta) (1 - \eta_C)}{(1 - \eta)^2}, & \text{for } \eta > 2 \eta_C - 1.
    \end{cases}
    \label{eq:constraint_xi_0}
\end{equation}
This bound is previously obtained as the universal upper limit of efficiency for arbitrarily given power \cite{Ryabov2016,Holubec2016,Shiraishi2016,Ma2018,Zhao2022,Ma2024}.

\begin{figure}
    \begin{center}
        \includegraphics{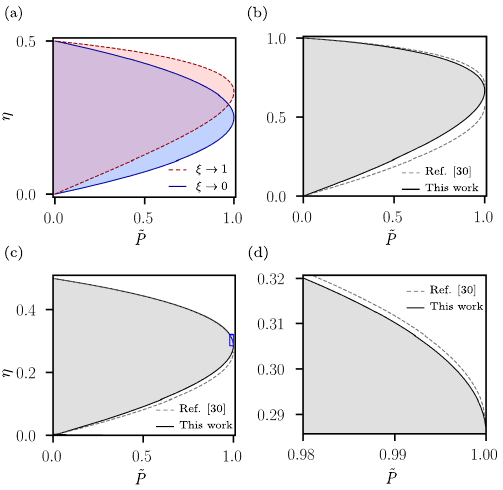}
        \caption{Exact power-efficiency constraint. The curves illustrate the analytical bound, the shaded regions illustrate the accessible regions for realistic engines. (a) Simplified constraint bounds with \(\xi \rightarrow 0\) (solid blue curves) and \(\xi \rightarrow 1\) (dashed red curves). The Carnot efficiency \(\eta_C\) is set as \(1/2\). (b) Constraint bound with \(\xi = 1/2\) and \(\eta_C \rightarrow 1\).  (c-d)  Constraint bound with \(\xi = 1/2\) and \(\eta_C = 1/2\). The axis range in panel (d) corresponds to the small blue box in (c). In FIGs. (b-d), the exact bound is shown with black solid curves, and the gray dashed curves indicate the previous bound in Ref. \cite{Ma2018}.
        }
        \label{fig:analytical_constraint}
    \end{center}
\end{figure}

In FIG. \ref{fig:analytical_constraint}(a), we show the power-efficiency constraint for both \(\xi \rightarrow 1\) (shown with blue solid curve) and  \(\xi \rightarrow 0\) (shown with red dashed curve). The Carnot efficiency is set as \(\eta_C = 1/2\). The corresponding shaded regions describe the accessible \(\tilde{P}\) and \(\eta\) for finite-time engines. From the above discussion, we conclude that our bound expressed with Eq. \eqref{eq:function_Delta} and TABLE. \ref{tab:Pi_constraint} can cover the previous results.

(3)  For intermediate scenarios with \(\xi = 1/2\) or \(m_c = m_h\), the dissipation in both high-temperature and low-temperature isothermal processes cannot be neglected. The constraint bound determined from Eq. \eqref{eq:function_Delta} becomes more complex. In this case, with \(\eta_C \rightarrow 1\), the bound is obtained as 
\begin{equation}
    \tilde{P}_m = \frac{2 (1 - \eta)}{\eta^2} \qty(27 - 18 \eta - \eta^2 
    - \sqrt{(1 - \eta)(9 - \eta)^3}
    ).
    \label{eq:xi_0.5_eta_C_1}
\end{equation}
In FIG. \ref{fig:analytical_constraint}(b), we plot the constraint bound in Eq. \eqref{eq:xi_0.5_eta_C_1} with black solid curve. For comparison, we also plot the approximated bound in Ref. \cite{Ma2018} with gray dashed curve. Here, the tight bound differs remarkably from the previous findings.

(4) For \(\xi = 1/2\) and \(\eta_C = 1/2\), the \(\tilde{P}_m\) is also obtained analytically from Eq. \eqref{eq:function_Delta} as
\begin{align}
    \tilde{P}_m = & 
    \frac{16 X}{3 \eta^2 ( 1 - \eta)} 
     - \frac{2^6}{3}
    \sqrt{
        \frac{X^2}{ 4 \eta^4 (1 - \eta)^2}
        +
        9\qty( 1/2 - \eta)
    }  \\
    & \times 
    \cos[(4\uppi+ \acos \Gamma ) / 3],
    \label{eq:xi_0.5_eta_C_0.5}
\end{align}
where \(X = 27/4 - 27 \eta + 33 \eta^2 - 12 \eta^3 - \eta^4\), and
\begin{align}
    & \Gamma = \qty[
        \frac{X^2}{4 \eta^4 ( 1 - \eta)^2}
        +
        9 (1/2 - \eta)
    ]^{-3/2} \\
    & \times \qty{
        \frac{27(1/2 - \eta)^2}{4 ( 1 - \eta)}
        -
        \frac{27 (1/2 - \eta) X}{4 \eta^2 ( 1- \eta)}
        -
        \frac{X^3}{8 \eta^6 ( 1 - \eta)^3}
    }
    .
\end{align}
In FIG. \ref{fig:analytical_constraint}(c-d), we plot the constraint in Eq. \eqref{eq:xi_0.5_eta_C_0.5} with black solid curve. The gray dashed curve is the bound in Ref. \cite{Ma2018}. The axis range in  FIG. \ref{fig:analytical_constraint}(d) corresponds to the small blue box in  FIG. \ref{fig:analytical_constraint}(c). It can be seen from FIG. \ref{fig:analytical_constraint}(c-d) that our result is obviously tighter than the previous results in Ref. \cite{Ma2018}.

\emph{Conclusion} -- In this letter, we have established the tightest power-efficiency constraint for thermodynamic cycles within the \(1/\tau\) entropy production model. Unlike previous results derived from approximations or restricted regimes, our bound is valid across the full parameter space, especially for large Carnot efficiency. Providing a benchmarking for evaluating the performance of heat engines.

\emph{Acknowledgements} -- 
We are grateful to H. Dong and Y. H. Ma for their helpful discussions. This work has been supported by the National Natural Science Foundation of China (NSFC) (Grant No. 12088101) and NSAF (Grant No. U2330401).

\bibliographystyle{apsrev4-2}
\bibliography{refs.bib}

\end{document}